\documentclass[prb,preprint]{revtex4-1}

\usepackage{amsmath}
\usepackage{amsfonts}
\usepackage{graphicx}
\usepackage{wrapfig}
\usepackage{subcaption}
\usepackage{natbib}
\begin{document}

\title{From Venus to the Parking Lot: A Flexible Stellar Interferometry Experiment}

\author{Eric J. Friedman}
\email{ejficsi@gmail.com}
\affiliation{International Computer Science Institute, Berkeley, CA}
\author{Michelle Goman}
\email{goman@sonoma.edu}
\affiliation{Sonoma State University, Sonoma, CA}

\date{\today}

\begin{abstract}
We describe a simple extension of our earlier Venus interferometry demonstration \citep{friedman2026simple} that can be performed indoors or outdoors, independent of planetary visibility or atmospheric seeing. A small flashlight is placed 15-50 meters away from  a telescope and covered with a circular aperture mask (hole diameter 2-6~mm). A small telescope fitted with an opaque mask bearing two 2~mm holes, spaced between 5 and 20~mm apart, is then used to observe interference fringes from this artificial extended source. By varying the hole separation and observing the fringes students can estimate the angular diameter of the source and verify their result directly by simple geometry, a check not possible with astronomical targets. The experiment requires no telescope tracking, no dark skies, and no special equipment beyond materials described in the original paper. It serves equally well as a pre-lab to build intuition before observing Venus, a standalone indoor daytime or outdoor nighttime activity, or a quantitative test of the interferometric method itself.

\end{abstract}

\maketitle

\section{Introduction}

In a previous paper \citep{friedman2026simple} we described how students can measure the angular diameter of Venus using nothing more than a small telescope and a piece of opaque plastic punched with two small holes. The demonstration is visually compelling and physically authentic - a miniature re-enactment of the landmark 1921 Michelson-Pease measurement of Betelgeuse \citep{michelson1921}. Its  limitation is obvious: Venus must be up, bright, far enough from the Sun to observe safely but before greatest elongation, a favorable geometry that recurs only about every 19 months in the evening. Even when the `stars align', clouds can spoil the party.

This extension was inspired by one such cloudy ``star night'' where fellow astronomers set up laptops with images of astronomical objects to look at through their telescopes, saving the event. We realized that this could be done for our interferometry experiment.

Here we describe our solution. We call it the ``parking lot'' version, though it works equally well in a long hallway, a darkened gymnasium, or any space where a light source can be placed 15-50 meters away from a small telescope. A small LED penlight, equipped with a simple aperture mask, becomes the artificial ``planet.'' As before, two-hole telescope mask is used to observe interference fringes and estimate the source's angular size. Because the geometry is fully under the experimenter's control, students can then \emph{verify} their interferometric measurement with a ruler and a bit of trigonometry; something that is impossible when the source is 170 million kilometers away.

The controlled-source version thus works at multiple levels of sophistication. Used before our previous lab, it builds intuition about fringes and what their disappearance means. Used after, it validates the technique. Used on its own, it is a self-contained, quantitative experiment that can be scheduled any night of the week or even daytime in a darkened corridor.

This experiment is a near duplicate of our previous one. We will describe it completely, but briefly. The reader is referred to our previous paper for more details, including intuition and background \cite{friedman2026simple}.

\section{The Experiment}

\subsection{Equipment}

\begin{figure}[htbp]
  \centering
  \begin{subfigure}[b]{0.48\textwidth}
    \centering
    \includegraphics[width=\textwidth]{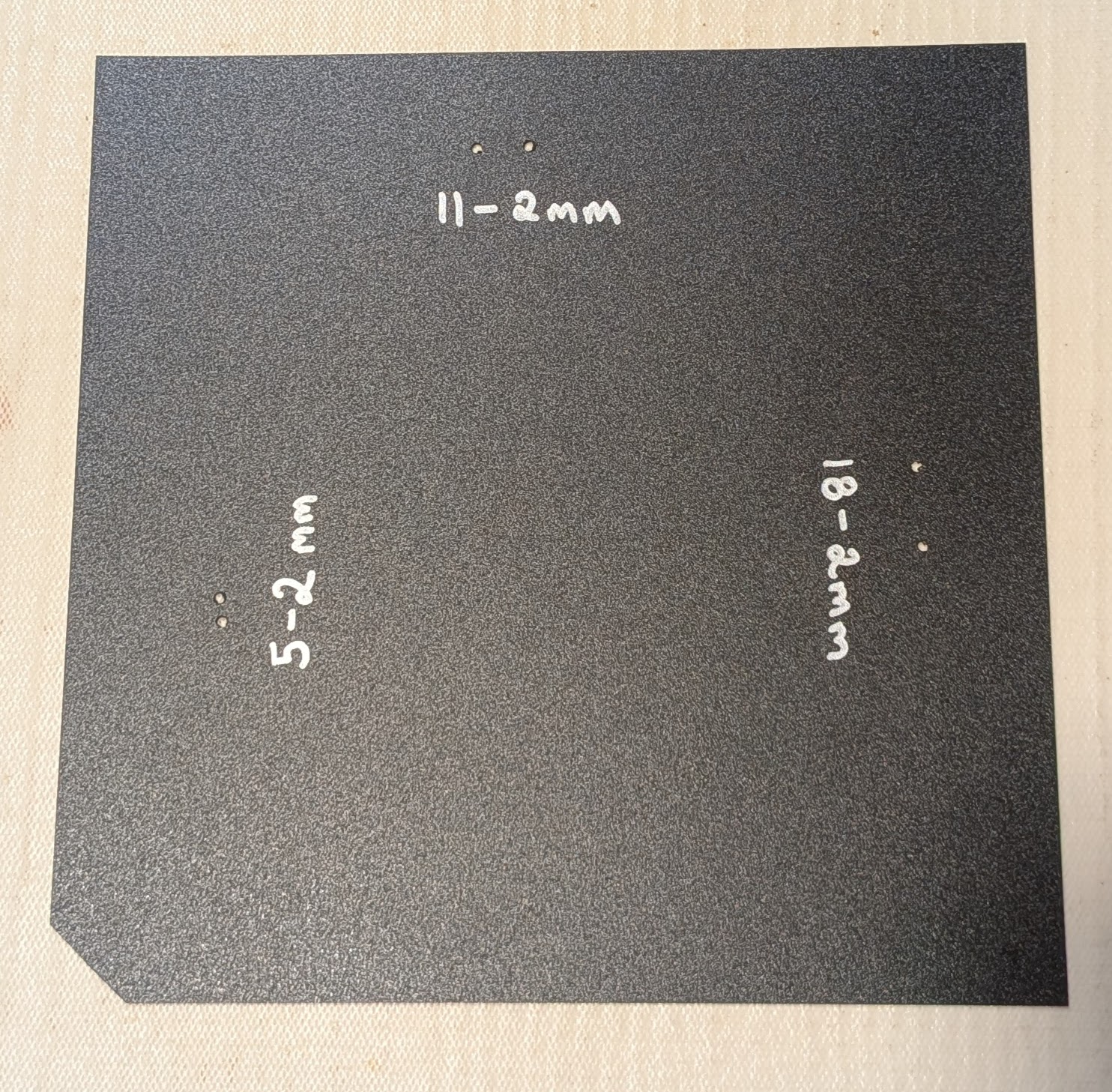}
    \caption{Interferometry Mask}
  \end{subfigure}
  \hfill
  \begin{subfigure}[b]{0.48\textwidth}
    \centering
    \includegraphics[width=\textwidth]{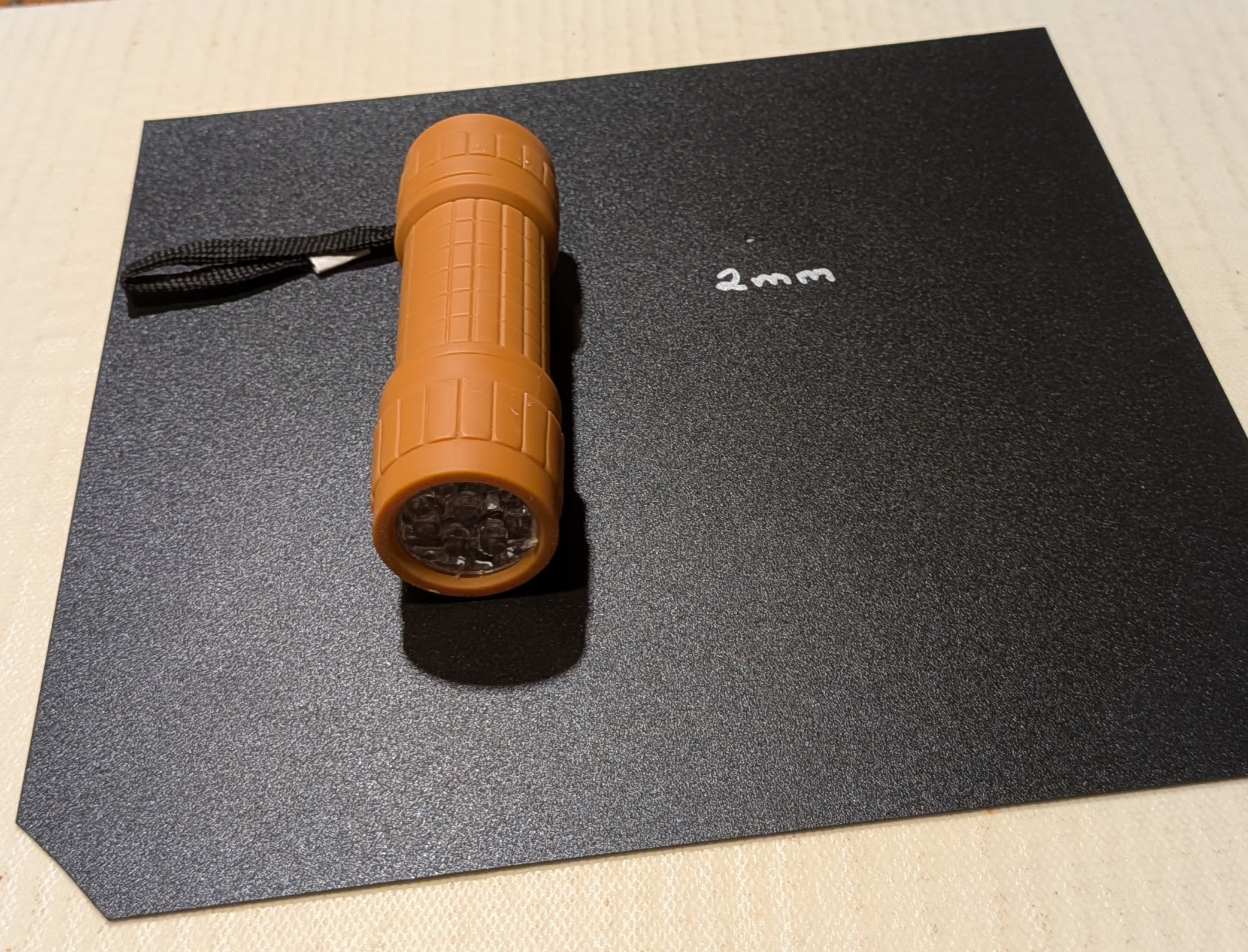}
    \caption{Source Mask and Flashlight}
  \end{subfigure}

  \caption{Masks and flashlight. Note that we put multiple interferometry masks on a single sheet and then cover the unused holes with masking tape or paper.}
  \label{fig:masks}
\end{figure}

{\bf Telescope}: Almost any planetary telescope with a long focal length ($>1000$ mm) and with a solid tripod will work. 

{\bf Interferometry and Source Mask}: Any opaque, thin but stiff material can be used. For this experiment, we have had success with 8”x 8” ``Black Opaque Rigid PVC Plastic Sheet'', 0.02 inch thick (available at most craft stores or online). Clean, well-defined holes are essential for the experiment. We used a power drill fitted with a 5/64-inch bit (a 2mm bit would be slightly more accurate), placing the plastic sheet over a piece of wood and drilling carefully to avoid tearing or deforming the holes. Hole spacings were varied from approximately 5 to 20 mm. Because the masks can be fabricated from simple, inexpensive materials, students can readily produce a range of configurations for experimental investigation (Fig.~\ref{fig:masks}). Note that the source mask must prevent bleed through while the interferometry masks is less critical; we have even used manila folders and even printer paper with pencil holes successfully for them. 

One might think that a diffuser of some sort would be necessary. However, we have found that the experiment  works well without the diffuser, while it is challenging to get sufficient brightness with a diffuser.
 
{\bf Flashlight}: Any small inexpensive flashlight should suffice.(Fig.~\ref{fig:masks}).

\subsection{Experimental Technique}

First, center the image within the eyepiece or camera view and focus carefully. Place the mask over the aperture and assess the visibility and sharpness of the resulting fringes. Repeat the measurement with masks of different spacings, identifying the mask size for which the fringes first become negligible or disappear entirely. Keep in mind that placing the mask over the aperture substantially reduces the image brightness.

There are many possible extensions of the experiment. (1) Students can count the observed fringes and compare the measured value with the corresponding theoretical prediction. (2) Students can try various hole arrangements with 3 or more holes in the Interferometry mask, trying to predict the patterns. (3) One can measure the fringe visibility in various ways and plot them against interferometry mask spacing. (4) Students can use multiple source hole geometries with 2 or more holes, to simulate binary  (or higher multiple) stars. The theory here is widely used and quite interesting \cite{argyle2012observing}. (5) Quantitative analyses estimating the accuracy of these measurements. (6) Interferometry masks with multiple holes in non-redundant geometry as used in modern astronomical satellites \cite{sivaramakrishnan2012non}.

Lastly we note that the parking lot version can be used in conjunction with the astronomical version in the original paper in a variety of educational ways.

\section{Analysis}

In this section we provide the basic formulas needed for the lab in the most relevant units. We refer the reader to our earlier paper \cite{friedman2026simple} for details, derivations and citations.

First we compute under the assumption that the source is a point. As in the previous analysis, all measurements are made using unfiltered white light, and we use $\lambda=550$ nm as a representative wavelength for the calculations.

Note that the size of the Airy disk in arcseconds as a function of interferometry mask hole diameter $D_I$ in mm is,
\begin{equation}
\theta_{\rm Airy}
= 2.44\frac{\lambda}{D_I}
\approx \frac{277}{D_I}
\quad\mathrm{arcsec}.
\end{equation}
Now the spacing of the fringes given a hole spacing of $W_I$ given in mm is 
\begin{equation}
\Delta\theta_{\rm fringe}
= \frac{\lambda}{W_I}
\approx \frac{113}{W_I}
\quad\mathrm{arcsec}.
\end{equation}
so the number of fringes on the Airy disk is given by
\begin{equation}
N_{\rm fringes}
\approx
\frac{\theta_{\rm Airy}}{\Delta\theta_{\rm fringe}}
= 2.44\frac{W_I}{D_I}.
\end{equation}

Now we take the finite size of the source into account.
Note that the angular diameter of the source in arcseconds  for a source mask hole diameter $D_S$ given in mm at a distance of $L$ m is 
\begin{equation}
\theta_{\mathrm{arcsec}}
= \frac{D_S}{L}\left(\frac{1}{1000}\right)
\left(\frac{180}{\pi}\right)(3600)
\approx 206\,\frac{D_S}{L}
\end{equation}
and by integrating over the source hole we see that the  fringe visibility is given by the function
\begin{equation}
V(W_I)
= \frac{2J_1(x)}{x},
\qquad
x=\frac{\pi W_I\theta_S}{\lambda},
\end{equation}
where $J_1$ is the first-order Bessel function and $\theta_S$ is the angular diameter of the source in radians. The first null in fringe visibility occurs when
\begin{equation}
x \approx 3.83,
\end{equation}
which gives
\begin{equation}
W_{I,\mathrm{null}}
= \frac{1.22\lambda}{\theta_S}.
\end{equation}

This provides a good approximation for when the fringes appear to vanish.

\subsection{Example}

As an example, take $L=32\,\mathrm{m}$ and a source hole $D_S=2\,\mathrm{mm}$. By Eq.~(4),
\begin{equation}
\theta_S \approx 206\,\frac{2}{32} \approx 13''.
\end{equation}
This matches a potential Venus angular diameter.

By Eq.~(7),
\begin{equation}
W_{I,\mathrm{null}} \approx \frac{138}{13} \approx 10.6\,\mathrm{mm}.
\end{equation}
Students testing masks from 5--20~mm should therefore see clear fringes near 5~mm and their disappearance near 11~mm, with very faint ones reappearing beyond this.

Taking an interferometer mask hole diameter $D_I=2\,\mathrm{mm}$ and the visible-fringe spacing $W_I=5\,\mathrm{mm}$, Eq.~(3) gives
\begin{equation}
N_{\rm fringes} \approx 2.44\,\frac{5}{2} \approx 6,
\end{equation}
fringes clearly resolved within the Airy disk at this spacing.

\section{Demonstration}

Using the masks and flashlight in Figure~1 we took photos using a Celestron 6SE telescope with a ZWO ASI676MC camera, a setup which is overkill for this experiment. 

The results are shown in Figure~2 below.

\begin{figure}[htbp]
  \centering
  \begin{subfigure}[b]{0.32\textwidth}
    \centering
    \includegraphics[width=\textwidth]{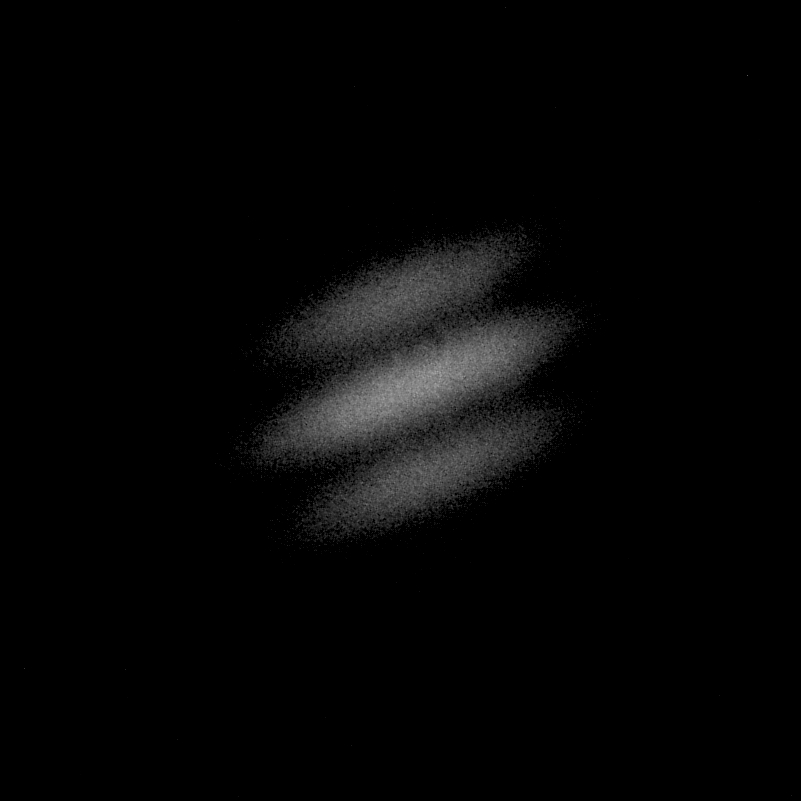}
    \caption{5 mm spacing mask}
  \end{subfigure}
  \hfill
  \begin{subfigure}[b]{0.32\textwidth}
    \centering
    \includegraphics[width=\textwidth]{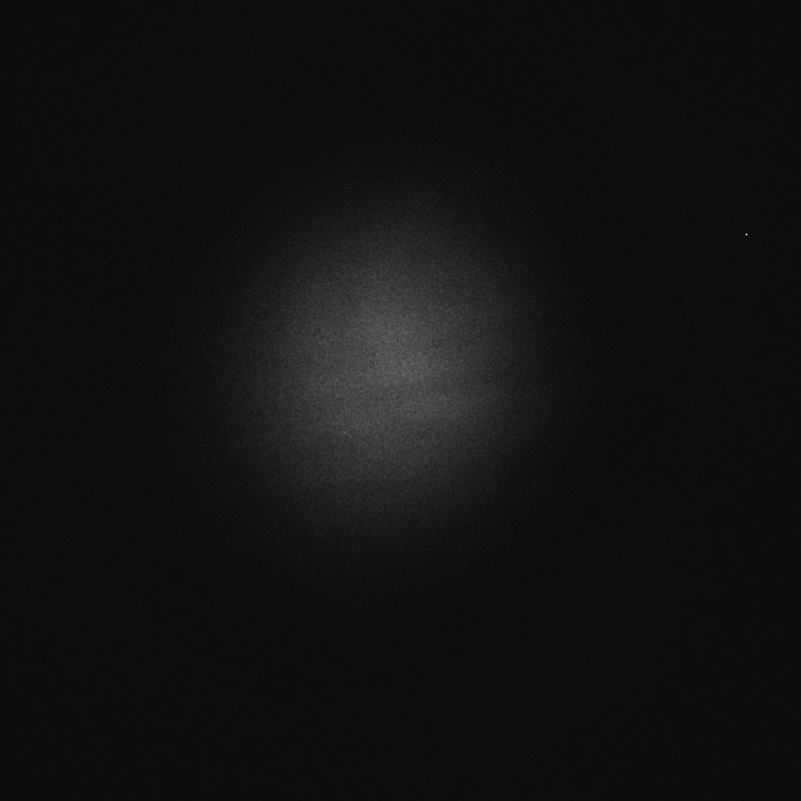}
    \caption{11 mm spacing mask}
  \end{subfigure}
  \hfill
  \begin{subfigure}[b]{0.32\textwidth}
    \centering
    \includegraphics[width=\textwidth]{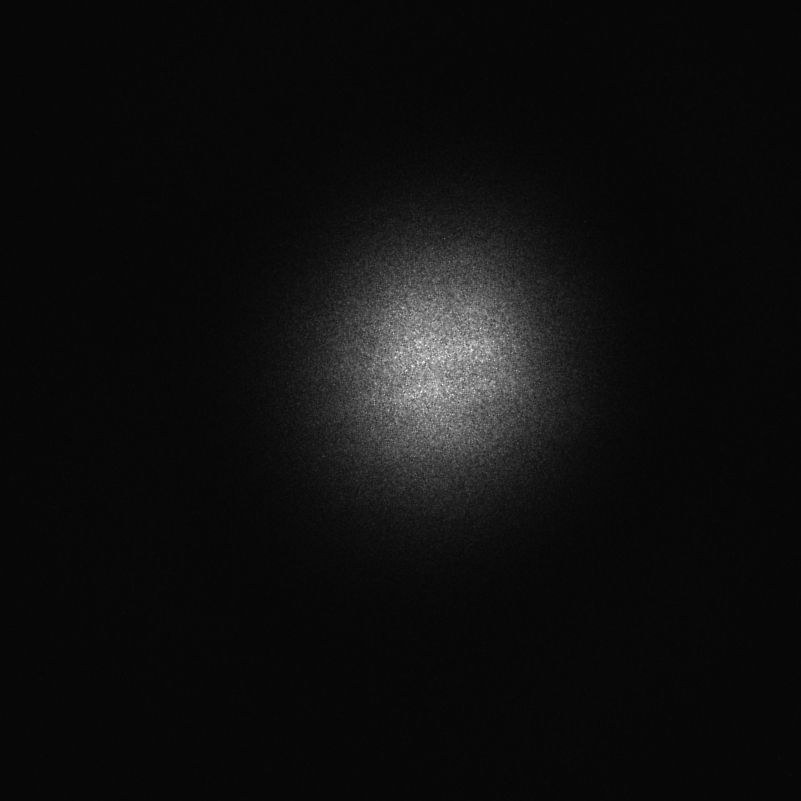}
    \caption{18 mm spacing mask}
  \end{subfigure}

  \caption{Demonstration with masks. Images are approximately 138'' in each dimension. See more discussion in the text.}
  \label{fig:demo}
\end{figure}

All holes in the following are 2mm in diameter.
Figure 2a shows the 2mm source with a 5mm spacing mask. Note that fringes are clearly visible. Fig 2b shows the lack of clear fringes near the first minimum, 11mm. Fig. 2c shows the slight fringing at 18mm nearing the second minimum. 

\bibliography{inter}

\end{document}